\documentclass[twocolumn,amsmath,amssymb]{revtex4}
\pdfoutput=1

\usepackage{natbib}
\usepackage{amsmath}
\usepackage{amssymb}
\usepackage{epsfig}
\usepackage{graphicx}
\usepackage{xcolor}
\usepackage{caption}
\usepackage{subfig}
\graphicspath{{figures/}}

\newcommand{\bi}{\begin{itemize}}
\newcommand{\ei}{\end{itemize}}

\begin{document}

\title{Searching for MHz Gravitational Waves from \\ Harmonic Sources}

\author{Jeronimo G.C. Martinez$^{1}$}\thanks{Corresponding author : jemartinezgc@gmail.com}
\author{Brittany Kamai$^2$}
\affiliation{\\ $^{1}$Department of Physics, University of Chicago, Chicago, USA;
$^{2}$Engineering \& Applied Sciences Division, California Institute of Technology, Pasadena, CA, USA; $^{2}$Department of Astronomy \& Astrophysics, University of California Santa Cruz, Santa Cruz, CA, USA}

\begin{abstract}

\noindent A MHz gravitational wave search for harmonic sources was conducted using a 704-hr dataset obtained from the Holometer, a pair of 40-meter power recycled Michelson interferometers. Our search was designed to look for cosmic string loops and eccentric black hole binaries in an entirely unexplored frequency range from 1 to 25 MHz. The measured cross-spectral density between both interferometers was used to perform four different searches. First, we search to identify any fundamental frequencies bins that have excess power above 5$\sigma$. Second, we reduce the per-bin threshold on any individual frequency bin by employing that a fundamental frequency and its harmonics all collectively lie above a threshold. We vary the number of harmonics searched over from $n= 4$ up to $n=23$. Third, we perform an agnostic approach to identify harmonic candidates that may have a single contaminated frequency bin or follow a power-law dependence. Lastly, we expand on the agnostic approach for individual candidates and search for a potential underlying population of harmonic sources. Each method was tested on the interferometer dataset, as well as a dark noise, photon shot-noise-limited, and simulated Gaussian-noise datasets. We conclude that these four different search methods did not find any candidate frequencies that would be consistent with harmonic sources. This work presents a new way of searching for gravitational wave candidates, which allowed us to survey a previously unexplored frequency range.

\end{abstract}
\pacs{??}

\maketitle

\section{Introduction}

Exploring the high frequency end of the gravitational wave spectrum is possible with a set of neighboring interferometers - the Holometer \citep{Holo2017_Instrument}. Initially, it was designed to look for a new kind of spacetime noise in the 1 to 25 MHz frequency range. Over 700 hours of data were collected with the Holometer to exclude predictions of quantum geometrical noise \citep{Holo2016_130hrs, Holo2017_700hrs}. We extend the utility of this dataset towards a gravitational wave search.

Before this search, the only constraints placed within the MHz range were from a cavity experiment \cite{Bernard2001} and a smaller Holometer dataset of 140 hours \cite{Holo2017_Kamai}. The first constraints came from a superconducting microwave cavity with optimal sensitivity at 1.38 MHz and a bandwidth of 100 Hz. Whereas, the Holometer dataset had a frequency range from 1-10 MHz and was two orders of magnitudes more sensitive than the cavity measurements. Now, we use a longer Holometer dataset with measurements between 1-25 MHz and sensitivity of $10^{-20} \ \rm{m/\sqrt{Hz}}$.

The kinds of searches that can be conducted with the Holometer are shaped by the fact that the time domain data was not stored; rather, the entire dataset is in the frequency-domain. This makes us insensitive to transient sources that other gravitational wave detectors search for. Given this feature, the previous Holometer search placed constraints on the stochastic gravitational wave background and primordial black hole binaries in circular orbits  \citep{Holo2017_Kamai}. 

In this paper, our analysis is designed to look for sources that can have harmonic signatures. There are two known sources that can emit gravitational waves with harmonics in this high frequency range: cosmic strings and primordial black hole binaries in eccentric orbits. Cosmic strings are one dimensional topological defects formed in the early universe. They arise from inhomogeneities in the spacetime distribution due to rapid and unstable phase transitions after the Big Bang \cite{DePies2007}. When these cosmic strings interact with each other, they form loops that emit gravitational waves at their fundamental and harmonic frequencies \cite{Kibble1976,Sanidas2012}. 

Cosmic string loops emit gravitational waves at their fundamental frequency and harmonics, defined as $f_n = n\times (2c/L)$ where $n$ is the harmonic mode, $L$ is the length of the string loop and $c$ is the speed of light. The gravitational wave power, $\rm{P}_{\rm{n}}$, emitted from cosmic strings is largest at the fundamental frequency and decays following a power law for subsequent harmonics, P$_{\rm{n}}$ $\propto$ n$^{-q}$. The exact value of $q$ depend on string formation, geometry, and evolution, with predicted $q$ values of $\approx$ 4/3 \cite{Sanidas2012, depies2009gravitational, Vilenkin:2000jqa}.

Searches for cosmic strings can target either individual sources or an unresolved stochastic background \cite{cosmic_string_1, cosmic_string_2, BLANCOPILLADO2018392, Nanograv11year_cosmic}. This search focuses on individual cosmic strings emitting at harmonic frequencies, given that previous constraints on the stochastic gravitational wave background were already conducted \cite{Holo2017_Kamai}. 

Primordial black holes are another early universe relic that could be emitting gravitational radiation at MHz frequencies. Primordial black holes binaries emit gravitational waves and those in non-circular orbits will emit in harmonics of their fundamental chirp frequency\cite{Peters1964}. The strength of their gravitational wave emission at harmonic frequencies is proportional to the eccentricity of their orbit \cite{Peters1964, Peters1963, Tanay_2016_eccentric, Moore_2016_eccentric, Hinder_2018_eccentric, Cao_2017_eccentric, Klein_2018_eccentric, Tiwari_2019_eccentric}. The previous Holometer search for primordial black holes binaries were for those in circular orbits. This new search allows for primordial black holes binaries in eccentric orbits. 

The approach presented here is an agnostic search of sources emitting gravitational waves with  harmonic emission patterns. We search over all frequencies between 1~to 25 MHz in the 704 hour dataset. To find evidence of any single source emitting gravitational waves, we vary the SNR threshold on each individual bin and number of harmonics. We use four different analysis criteria and compare the results against noise datasets and simulated datasets. Below, we describe the instrument, dataset, analysis pipeline, and conclusions.


\section{Instrument}\label{sec:instrument}

The Holometer is comprised of two identical power-recycled Michelson interferometers operated at the Fermi National Accelerator Laboratory \citep{Holo2017_Instrument}. Each 40 meter interferometer is housed within separate ultra-high vacuum systems and equipped with separate lasers, injection optics, electronics, and core optics (beamsplitter, power-recycling mirror, and two end mirrors). These interferometers are nested within the same spatial orientation and separated by 0.5 meter.

With this design, cross-correlation techniques allow us to reach sub-shot-noise length sensitivity as compared to a single interferometer. The dominant noise source at MHz frequencies is photon shot noise and each interferometer is operated with independent lasers. Therefore, the length sensitivity increases as $1/\sqrt{N}$, where $N$ is the number of Fast Fourier Transforms (FFTs).

The Holometer was initially designed to search for broadband noise, which resulted in a choice to store only the frequency domain data \citep{Holo2017_Instrument}. The saved data is the power spectral density of the individual interferometers, the cross-spectral density between the interferometers, and cross-spectral densities of the interferometer output with the environmental monitors. This stands in contrast with the lower frequency experiments such as ground-based detectors such as LIGO, Virgo, KAGRA, and pulsar timing arrays such as NanoGrav, PPTA, EPTA,  \citep{2015_Advanced_LIGO, VIRGO2015, amaroseoane2017laser_LISA, kerr2020parkes, GEO600_2004, EPTA2015, PPTA2010, 2018ApJS..235...37A, PhysRevD.88.043007, LIGO2016_InstrumentLong, GW190425} that store all of their time series data.

A brief summary of the data acquisition procedure is as follows - The output of each interferometer is sampled at 50 MHz and after $\sim$3-milliseconds, the time series data is fast-Fourier transformed. The real-valued, power spectral density (PSD) is calculated for each interferometer : $\rm{PSD} = |A|^2$, where $A$ is the amplitude of the Fourier transform. Also, the complex-valued, cross-spectral density (CSD) is computed between the output of each interferometer : CSD is $A_1 A_2 e^{i(\theta_1 - \theta_2)}$ where 1 and 2 corresponds to each interferometer and $\theta$ is the angle of the vector in the complex plane for interferometer $1$ and $2$.  After 1.4 seconds, all of the computed millisecond power- and cross-spectra are averaged together which is then GPS-time-stamped and recorded. Data vetoes were implemented to ensure that contaminated data (i.e. from large radio-frequency interference spikes) were not included in the averaging. This procedure repeats throughout the extent of the observing run. This procedure is described in more detail in \citep{Holo2017_Instrument}.

To verify whether the data acquisition pipeline would be able to recover correlated signals, we placed LEDs in front of the detectors of each interferometer. We sent a 13 MHz signal into the LEDs and monitored this signal throughout the science runs.

Extensive campaigns were conducted to verify that there are no unknown correlated noise sources above 1 MHz.  Noise sources were characterized through measurements taken before, during and after observing runs. These environmental monitors were used to quantify the amount of contamination from radio-frequency interference, laser phase and intensity noise. Additional details on each of these studies are included in the following references \citep{Holo2017_Instrument,Lanza2015,McCuller2015,Kamai2016,Richardson2016}). 


\section{Datasets}\label{sec:datasets}
We employ the use of three datasets to search for harmonic sources and validate our analysis algorithms. The main dataset is 704 hours of interferometer data that has the real-valued, power-spectral densities for each interferometer and the complex-valued, cross-spectral densities between the two interferometers. The second dataset is 320 hours of lightbulb data, which is an independent noise source used to replicate the high photo-current seen by the detector during standard operations. The third dataset is 13 hours of dark noise data that preserved the same RF environment and measured the signal detector response with no light on it. Below includes more details about each of these datasets.

\subsection{Interferometer Dataset}\label{sec:ifodataset}

The main dataset used in this search is the integrated 704 hours of interferometer running time obtained between the months of July 2015 and April 2016.

\begin{figure*}

  {\includegraphics[width=\textwidth]{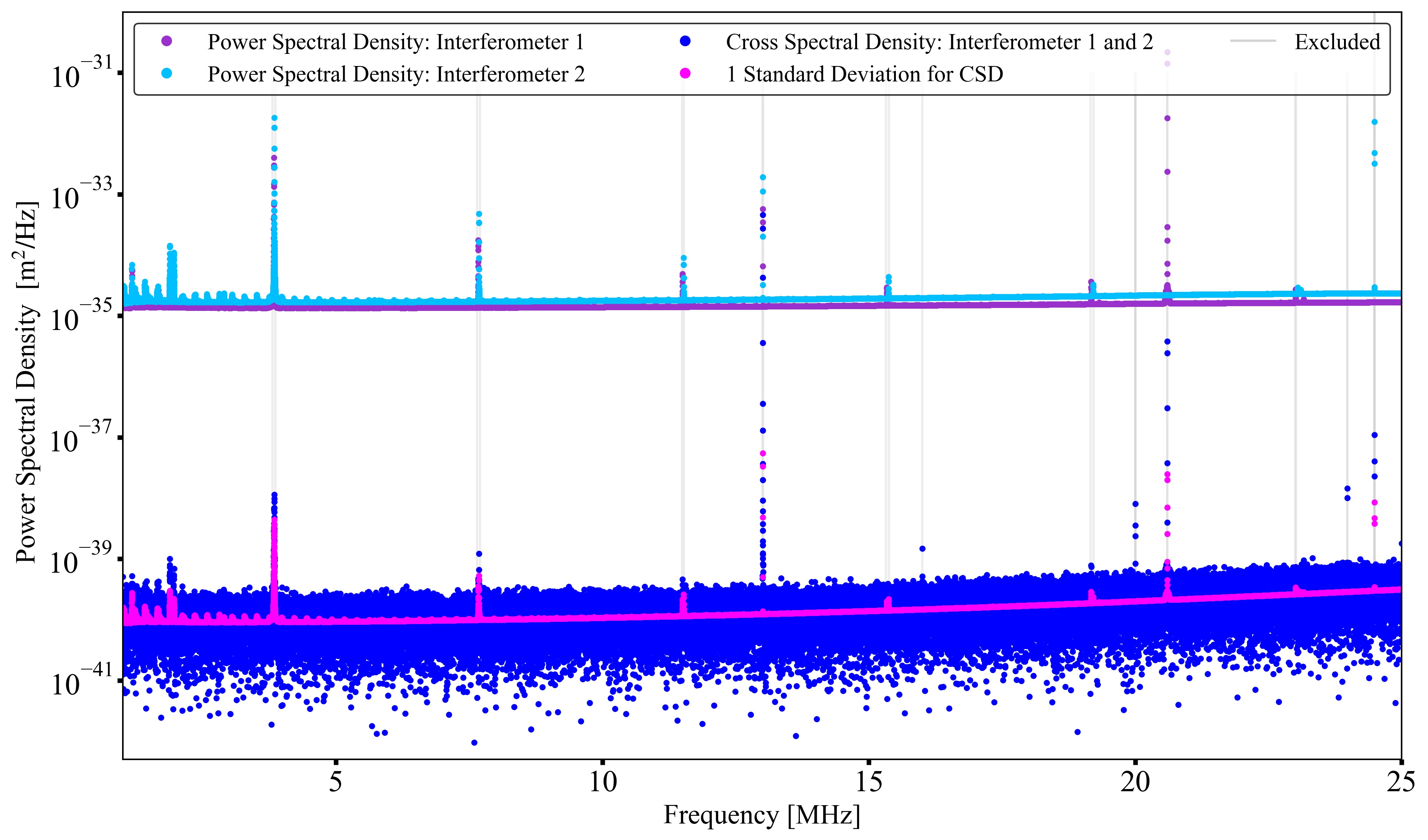}}
  \caption{The magnitude of the power spectral density of the 704-hr Holometer dataset used throughout this study. The Holometer consists of two power-recycled Michelson interferometers, which were operated with stored cavity powers of 2 kW. The light blue and purple traces represent the integrated precision for each of the respective interferometers  throughout the entire observing run. Given the integration time and the power in each interferometer, the pink trace represents the expected cross-spectral density for two uncorrelated noise sources. The dark blue points is the measured cross-correlation between the two interferometers. Across a wide range of frequencies, the measured CSD is consistent with what is expected given the integration time and power levels. This study is designed to look closer at each individual frequency bin in the search for harmonic sources. We excluded peaks (highlighted in grey boxes) that are known sources of noise within the instrument - the peak at 3.83 MHz (and its corresponding harmonics) is the free-spectral range of the 40m Fabry-Perot Interferometer, the peak at 13 MHz was injected for calibration purposes of the readout system and is generated from an LED placed directly in front of the signal detectors, and the two peaks at 20.5 and 23.5 MHz are used to phase lock the lasers to the resonant interferometer cavities. }\label{fig:full_ifo_data}
\end{figure*}

The averaged dataset is shown in Figure~\ref{fig:full_ifo_data} where the PSD from each of the interferometers are shown in light blue and purple, respectively. The magnitude of the CSD between the two interferometers is shown in dark blue. The pink trace is the standard deviation noise line on the CSD measurement.

An intuitive way of understanding the variance on the CSD measurement is based on the noise model of the experiment. Given that we have two uncorrelated noise sources (photon shot noise in separate interferometers), the expected level of the CSD given no correlations would be $\sqrt{\rm{PSD}_1 \times \rm{PSD}_2/\rm{N}}$, where N is the number of FFTs. In practice, the CSD variance is calculated during each timeframe from the individual 1,400 millisecond CSD measurements. More details of how this was calculated are in \citep{Holo2017_Instrument,McCuller2015}

The frequency range evaluated in this study is from 1 to 25 MHz. The lower frequency cutoff comes from avoiding a region where correlated laser noise has been identified in previous studies \citep{Holo2016_130hrs} whereas the upper frequency cutoff is due to the sampling rate. The frequency resolution is 382 Hz and the number of frequency bins within this range is 62,914. 

\subsection{Lightbulb Datasets}\label{sec:lightbulbdata}
The second dataset is the lightbulb dataset that was generated to simulate standard operating conditions by having the same photon shot noise levels on each signal detector. Rather than having power from the laser enter the instrument, the detectors measured a signal which came from independent bright incandescent lightbulbs placed directly in front of each signal detector. The lightbulb dataset contains the power spectral density of each interferometer and their cross-spectral densities for 320 hours of integration. For the data collection, the laser was shuttered (meaning there was no laser light entering the interferometer) and all the electronics were left running in their nominal operating configuration. The lightbulbs provide an uncorrelated Gaussian noise signal across both interferometers. This dataset is run through the analysis in Sections~\ref{sec:harmonic_constant}, ~\ref{sec:harmonic_unconstrained}, ~\ref{sec:harmonic_population} and we compare these results against the results from the interferometer dataset.

\subsection{Dark Noise Dataset}\label{sec:darknoisedata}
The third dataset is the dark noise dataset, which is used to quantify the ambient electronic environment. This dataset contains the cross spectral density between the interferometer signal detectors and represents the amount of noise in the signal detectors, digitizers and any additional contamination coming from the electronics that is runs to the interferometers. During the 13 hours of integration, the laser was shuttered and all the electronics were left running in their nominal operating configuration. Without any light source in the interferometers, this dataset is not shot-noise limited. Therefore, this data is only used if we need to compare harmonic source candidates against background ambient noise.


\section{Analysis}~\label{sec:analysis}
We construct an analysis pipeline to identify sets of frequency bins with excess power (in the cross-spectral density measurement) that are consistent with a potential source of interest emitting at its fundamental and harmonics. If a gravitational wave signal were to interact with the Holometer, it would appear in phase in both interferometers. This is because the two interferometers are co-located, separated by less than 1 meter and we are evaluating the signal in the MHz frequency band. Therefore, our search is to look for candidates in the positive, real component of the CSD. 

Throughout this entire analysis, we exclude frequency bins that have known origins. These frequencies are 3.8 (and its harmonics), 13 (and its first harmonic), 20.5 and 24.5 MHz bins. Each of these center frequencies have a range of bins that on either side of the peak that are also excluded. One thing to note is that some of these frequencies that lie above the 25 MHz range are folded back to lower frequencies due to Nyquist sampling. For example, the 13 MHz harmonic lies at 24 MHz rather than the expected 26. A similar folding happens for the 20.5 and 24.5 frequencies that are used to drive each individual laser.  Table~\ref{tab:frequencies} enumerates the contaminated frequency bins and their identified origins. Additionally, all excluded bins are highlighted in Figure~\ref{fig:full_ifo_data}.

To search for any candidate sources, we calculate the z-score for each frequency bin. The z-score is calculated as the ratio between the cross-spectral density (CSD) and the standard deviation for that cross-spectral density($\sigma_{\rm{CSD}}$) which are generated during the data collection described in Section~\ref{sec:instrument}. Explicitly,

\begin{equation}
{\rm Z\,\,\,score}(f) = \frac{{\rm CSD}(f)}{ \sigma_{\rm{CSD}}(f)}
\label{eqn:zscore}
\end{equation} 

\noindent where $f$ is frequency. Z-scores are calculated for each dataset described in  Section~\ref{sec:datasets}. We use this to flag any potential frequency
bins that require further follow-up when calculated for the interferometer dataset.

To search for harmonic sources, we apply various thresholds to the  z-score values to identify potential candidates. A harmonic source is defined as a single fundamental frequency and its $n$-harmonics whose emission signal pattern can be identified against noise.

We perform four different types of searches to identify candidates. First, we perform an excess power search to identify any fundamental frequencies that have a z-score value above a given threshold. Those frequencies are used as harmonic source candidates, which we follow up by examining the z-score of their harmonic frequencies for evidence of excess power. Second, we construct a search that requires the z-score of a fundamental frequency and its harmonics to all lie above a threshold value. This allows us to reduce the z-score threshold at any individual frequency bin while maintaining a high statistical significance for detecting a harmonic source.

Third, we perform an agnostic approach to identify harmonic candidates that may have a single contaminated frequency bins or follow a specific power-law dependence. In this test, we assign each z-score its corresponding probability from a Gaussian distribution. These probabilities are multiplied together and we use thresholds on these products to identify individual harmonic source candidates. Lastly, we expand on the agnostic approach for individual candidates and search for a potential underlying population of harmonic sources. We search for this population within the positive-real quadrant of the interferometer dataset. We compare that to simulated datasets from pure Gaussian noise and Gaussian noise with an injected underlying population of harmonic sources. Additionally, we compare this distribution to the other quadrants of the interferometer dataset and lightbulb dataset. 


\begin{figure*}
  \includegraphics[width= 0.90\textwidth]{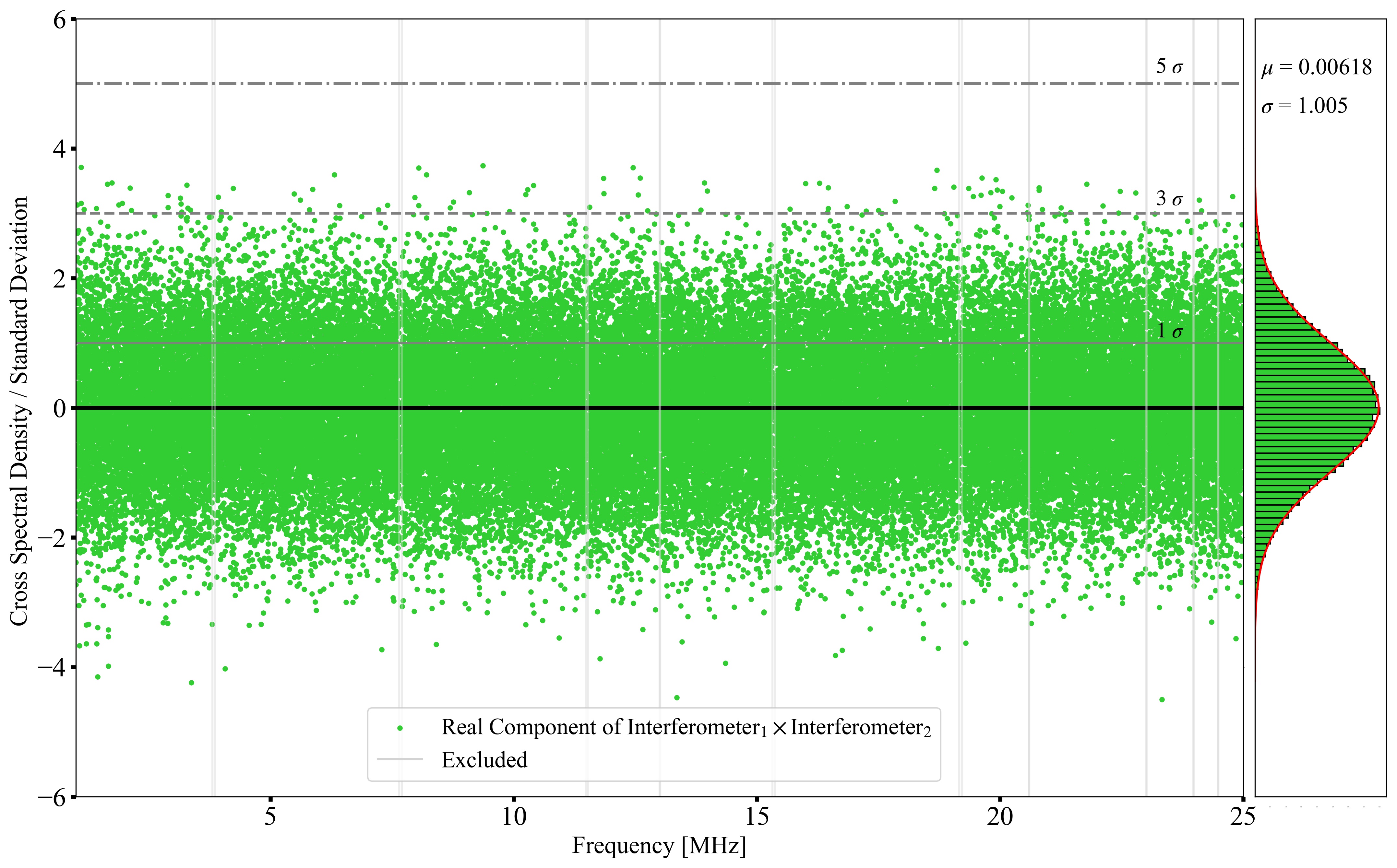} 
   \includegraphics[width=0.90\textwidth]{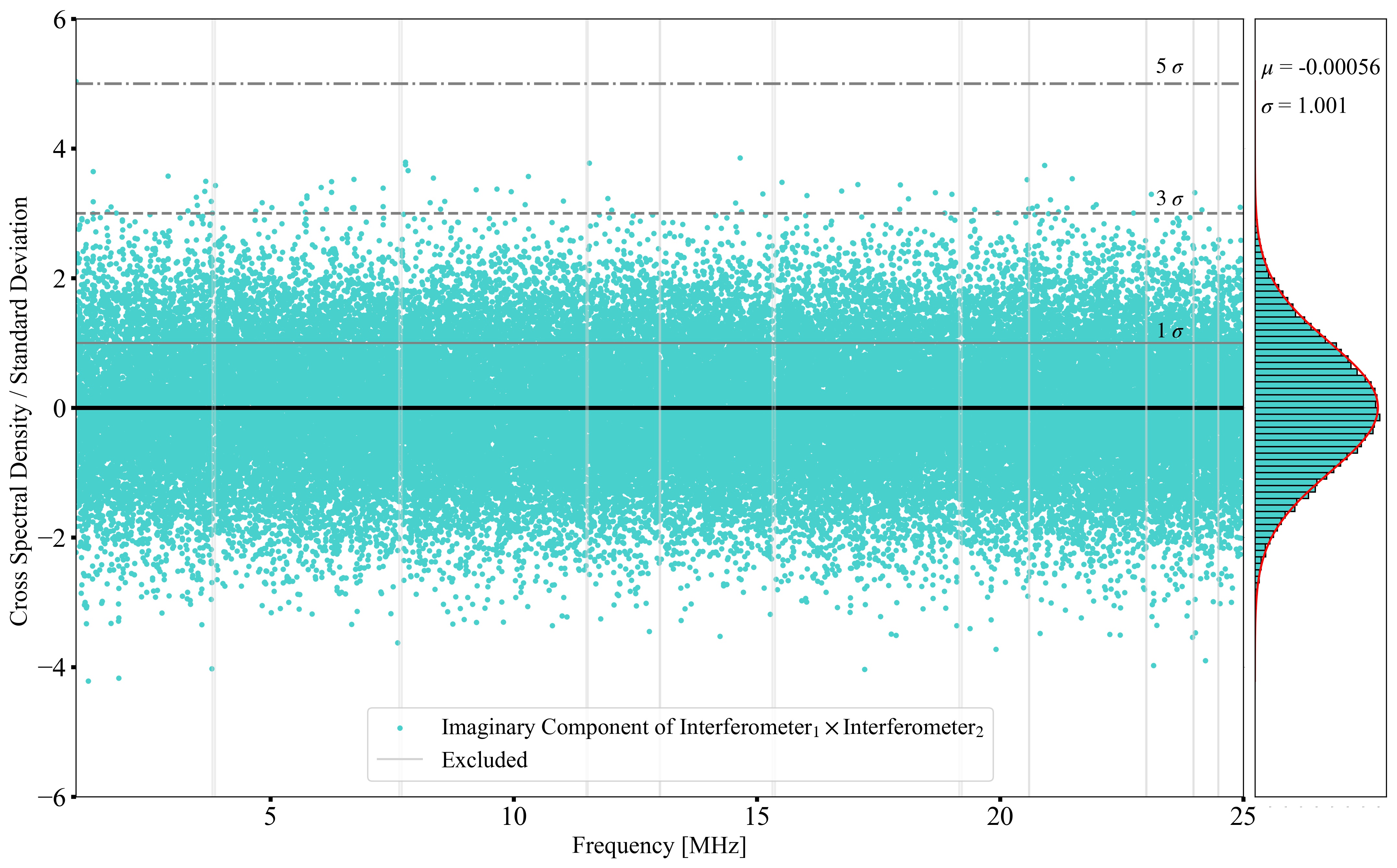}
\caption{{The calculated z-score for each component of the cross-spectral density measurement as a function of frequency. The top plot is of the real component of the CSD and the bottom plot is the imaginary component of the CSD. The vertical histograms show the statistical distribution of the data along with their mean and standard deviation statistics. The noise follows a Gaussian distribution, which is used to map the z-score values onto their corresponding probability for each individual frequency bin. We included 1, 3, and 5$\sigma$ lines for reference as well as excluded frequency ranges which we know to be contaminated with additional noise and are explained in Table ~\ref{tab:frequencies}. If a gravitational wave signal interacted with the Holometer, it would appear as an in-phase signal between both interferometers. Therefore, the analysis conducted searches for harmonic sources candidates within the positive-real CSD quadrant. The negative-real, positive-imaginary, and negative-imaginary data quadrants were used to verify our search algorithms and compare to the results found in the positive-real quadrant. In Section~\ref{sec:excesspower}, we look for individual frequency bins with statistically significant z-scores in the positive real quadrant. Our search is to identify any frequency bins above 5$\sigma$ thresholds. Across the 1 to 25 MHz range, we do not find any single frequency bins with significant emission. This extends the range of excluded narrow-lined gravitational wave emitters from 1.92 MHz~\cite{Holo2017_Kamai} up to 25MHz.  \label{fig:ifoCSDdistributions}
}
}
\end{figure*}

\begin{table}
	\title{\bf{}}
	\resizebox{\columnwidth}{!}{
		\medskip
		\begin{tabular}{| c | c | }
			\hline
			Frequency [MHz]   &Source \\ \hline
			3.83 $\pm$ 25 kHz (\& harmonics)  & Free-spectral range of Interferometer \\ \hline
			13 $\pm$ 8 kHz   &LED signal for calibration  \\ \hline
			20.5 $\pm$ 2 kHz &Used for laser stabilization \\ \hline
			24.5 MHz $\pm$ 2 kHz &Used for laser stabilization \\ \hline
		\end{tabular}
	}
	\caption{Frequency bins excluded from this analysis and their known sources. } \label{tab:frequencies}
\end{table}

\subsection{Excess Power Search}\label{sec:excesspower}

An approach to search for narrow-lined sources is to search for excess power in a single frequency bin. Figure~\ref{fig:ifoCSDdistributions} shows the calculated z-score values for each individual bin from 1 to 25 MHz for real and imaginary components of the interferometer dataset. The vertical figures shows that the real and imaginary components follow a Gaussian distribution. 

In our excess power search, we look for candidates in the positive-real quadrant of the interferometer dataset.  Figure~\ref{fig:ifoCSDdistributions} shows all four quadrants of the interferometer dataset with 1, 3, and 5$\sigma$ lines to guide the reader. In a previous search done at MHz frequencies, the Holometer collaboration argued that given the dominant noise sources at MHz frequencies is photon shot noise\citep{Kamai2016, Holo2017_Kamai}, then an allowable threshold for an excess power can be set to 5$\sigma$. In this earlier work, an excess power search was done up to 1.92 MHz using the first 140 hours of data that was available\citep{Holo2017_Kamai}.

Our work extends on this search by investigating beyond 1.92 MHz up to 25 MHz. Additionally, our dataset has better sensitivity given the 704 hour acquisition time. Even with this additional sensitivity and extended frequency range, we do not identify any individual frequency bins with z-scores above 5$\sigma$. We conclude that this search criteria is too stringent to search for any harmonic candidates.


\subsection{Harmonic Search: Algorithm }\label{sec:algorithm}

 Given that we are searching for harmonic sources, we can incorporate this into our algorithm and create a new kind of search that accounts for this feature. To search for gravitational wave emitting source candidates, we require that a fundamental frequency and all of its harmonics have excess power in each of their frequency bins.  This allows us to reduce the per-bin threshold that was previously set in the excess power search (Section \ref{sec:excesspower}). 

Here we describe what is calculated for any generic CSD dataset:
\begin{enumerate}
	\item Calculate the z-score for each frequency bin using Equation~\ref{eqn:zscore}.
	\item Assign the corresponding probability to each z-score values for a Gaussian distribution (described in detail below).
	\item Choose the number of harmonics to search over, $n$.
	\item Designate the first frequency bin to be the fundamental frequency.
	\item Calculate the ``combined probability" by multiplying the probabilities of the fundamental frequency and its harmonics using Equation~\ref{eqn:combinedprob}.
	\item Repeat 3 \& 4 through the entire list of frequencies until there are no longer $n$-harmonics available in the frequency band.
	\item If the  ``combined probability" values are less than the designated threshold value, then return the list of fundamental frequencies, probabilities, z-score and CSD values for further evaluation.
\end{enumerate}

Throughout the analysis, we utilize both the z-score value and its associated probability. The z-score value is a more intuitive representation of how far above the noise a particular CSD measurement is.  However, when we calculate the combined probability of each fundamental frequency and harmonics, we use probability values rather than the z-score.

 As seen in Figure~\ref{fig:ifoCSDdistributions}, the interferometer dataset follows a Gaussian distribution and therefore we can map z-scores to their corresponding probability using a one-tailed statistics test: the assigned probability is the integral of the Gaussian from the z-score value to infinity. This assignment of probabilities is done based on the dataset and the quadrant (positive-real, negative-real, positive-imaginary, negative-imaginary) that is being tested. 

Given that the harmonics correspond to the same source, we can multiply the probability of each individual harmonic frequency together to get a total combined probability. This criteria allows us the capability
to reduce the signal-to-noise threshold on any individual bin.

For each fundamental frequency and  $n$-harmonics, a combined probability is calculated using:
\begin{equation}
\rm{Probability}_{\rm{combined}} = \prod_{i = 0}^{n} \rm{Probability}_{\rm{i}} \label{eqn:combinedprob}
\end{equation}

where i is the harmonic number, and Probability$_{\rm{i}}$ is the probability associated with the i'th harmonic's z-score assuming a Gaussian centered about zero. 

~~~~~~~~~~~

\subsection{Harmonic Search: Minimum Threshold}\label{sec:harmonic_constant}

In the following harmonic tests, we apply thresholds to the z-scores of the fundamental and harmonic frequencies. To identify harmonic source candidates, we require that all of the z-scores of a fundamental frequency and harmonics are above the assigned thresholds. We first perform a constant threshold test which assigns the same threshold for all frequencies of the harmonics. Second, we preform a power-law dependence threshold test where the assigned threshold decreases following a power law for higher harmonic frequencies. Any identified fundamental frequencies and harmonics whose z-scores are above the thresholds are returned as harmonic source candidates for follow-up. 

Section~\ref{sec:constant_threshold} details the constant threshold test where the same z-score threshold is assigned for all frequencies in a set of fundamental and harmonic frequencies. This is a conservative approach that does not assume a particular emission model and imposes a requirement that all frequencies will emit gravitational waves with a power that is higher than the noise. 

In section~\ref{sec:power_law}, we introduce a power-law dependence into the threshold test. We use the same framework introduced in the the constant threshold except that the z-score threshold decreases according to a power law with increasing harmonics. This test is done to be more sensitive to conventional harmonic sources like cosmic strings that are expected to emit gravitational waves with a power law dependence.  

\subsubsection{Constant Threshold}\label{sec:constant_threshold} 

In this first threshold test, we set the $\rm{Probability}_{\rm{combined}}$ to a fixed value and require that all of the individual probabilities are below a minimum constant probability threshold. Throughout this part of the analysis, we test different combinations of the number of harmonics and apply various thresholds for what the combined probability value can be. This means that the per-bin probability threshold needs to be calculated for each combination. Using equation~\ref{eqn:combinedprob}, we calculate the necessary per-bin probabilities using the following relationship:
\begin{equation}
\rm{Probability}_{\rm{individual}}^{{\rm const.\,thres.}} = \sqrt[\leftroot{-2}\uproot{2}n + 1]{\rm{Probability_{combined}}} 
\end{equation}

where the $\rm{Probability}_{\rm{individual}}^{{\rm const.\,thres.}}$ is the probability threshold that each harmonic must be below and $n$ is the number of harmonics of the fundamental frequency.

\begin{figure*}
  \centering 
  {\includegraphics[width=0.9\textwidth]{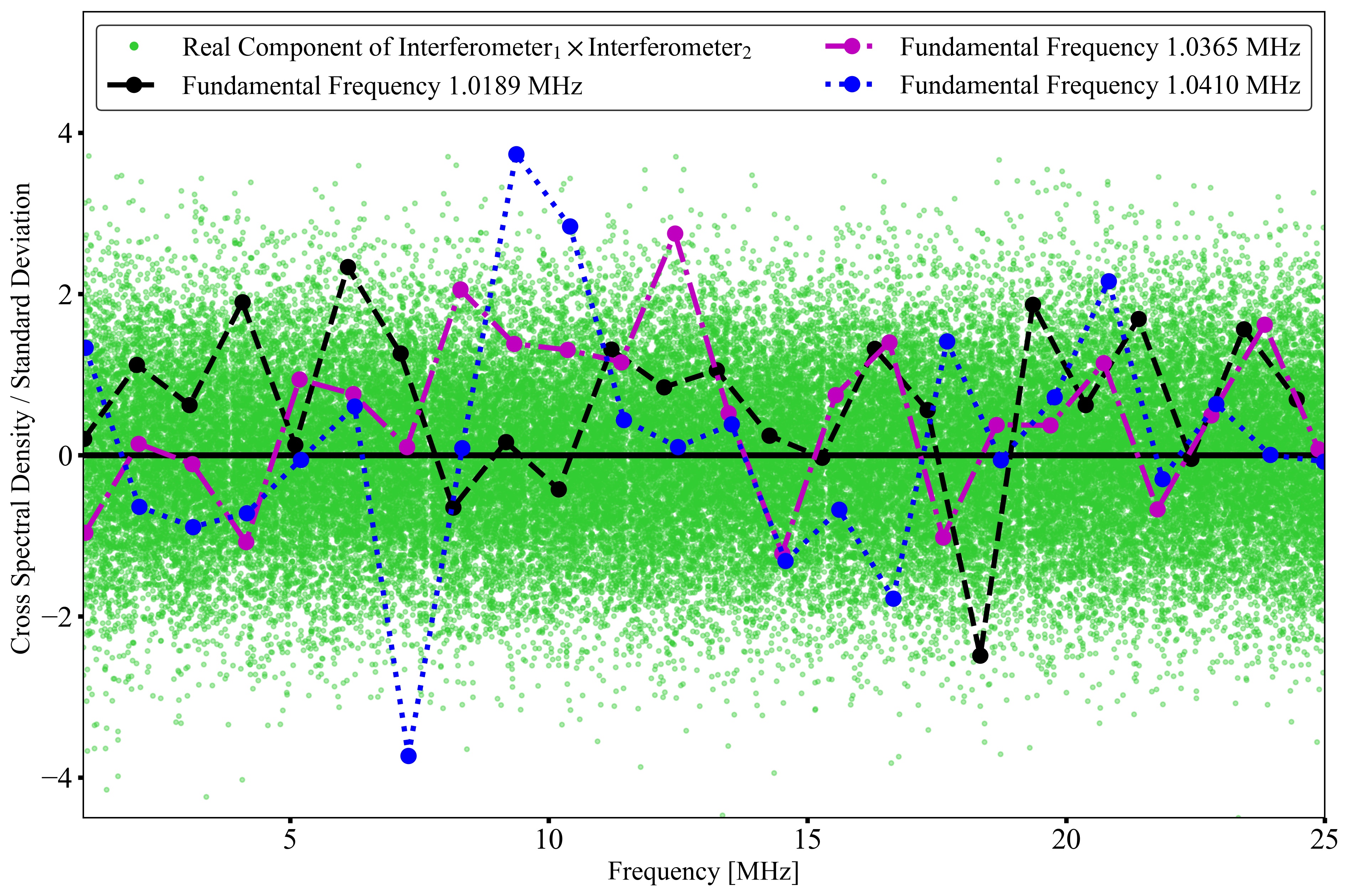}}
  \caption{The real component of the interferometer CSD dataset as a function of frequency with three highlighted fundamental frequencies and harmonics identified in Section~\ref{sec:harmonic_unconstrained}. In that search, we ran an agnostic test where we allowed the z-scores to have any value. Each z-score was converted to its associated Gaussian probability and the z-scores for a fundamental frequency and harmonics were multiplied together to generate a combined probability using Equation~\ref{eqn:combinedprob}. The identified fundamental frequencies in this figure (black, blue, and magenta) are the three fundamental frequencies with the lowest combined probabilities identified in the positive-real quadrant of the interferometer dataset. The unconstrained search allowed for the possibility that single frequency bins might contain additional noise that would be excluded from the thresholds tests. If any of these frequencies were to follow a harmonic source pattern, we anticipate there to be either excess power in most frequency bins or for the z-scores to follow an inverse power law for harmonic frequencies. Additionally, we expect that the emission pattern and combined probability values would differ from those found in the other three dataset quadrants (negative-real, positive-imaginary, and negative-imaginary) and the lightbulb dataset quadrants. None of the highlighted frequencies are consistent with these criteria and  we conclude there is no supporting evidence for gravitational wave candidates using this method.}\label{fig:harmonicTests}

\end{figure*}

 As an example of how this process is done, for a test of 9 harmonics, we start by setting $\rm{Probability}_{\rm{combined}}$ equal to 2.9$\times 10 ^{-7}$, which corresponds to the equivalent 5$\sigma$ of a Gaussian distribution. With these conditions, each fundamental frequency and all 9 harmonics must have z-scores with an equivalent probability that is below 0.22. A probability of 0.22 corresponds with a z-score being above 0.77$\sigma$ and therefore we set 0.77$\sigma$ as our z-score constant threshold for this test.

We conduct different tests for harmonic numbers ranging from 4 to 23. Starting with the first frequency in the array, the z-score of the fundamental frequency and all 4 harmonics are compared to the z-score threshold. After this, the algorithm checks the next fundamental frequency in the list until running out of fundamental frequencies whose 4th harmonic still lies within the 25 MHz frequency limit. Any fundamental frequencies where all harmonics lie above the z-score threshold are returned as candidates of interest. Then the search is started again for 5 harmonics and continues until testing through 23 harmonics.

Using this search algorithm, we ran our tests for various combinations of harmonics that had the equivalent $\rm{Probability}_{\rm{combined}}$ values set to 2.9 $\times 10^{-7}$, 9.9 $\times 10^{-10}$, 1.3 $\times 10^{-12}$, 5.6 $\times 10^{-16}$, which corresponds to 5, 6, 7 and 8$\sigma$ of a Gaussian distribution. We found that there were no fundamental frequencies and harmonics where all were above the z-score threshold. When we reduced the combined probability to 3.2 $\times 10^{-5}$ , which corresponds to 4$\sigma$, we recovered one fundamental frequency candidate. We ran these same condition of setting the $\rm{Probability}_{\rm{combined}}$ to 3.2 $\times 10^{-5}$ on the negative-real, positive-imaginary and negative-imaginary quadrant and recovered a similar number of candidates. After comparing the z-score as a function of frequency for the identified candidates of each quadrant, we found that their patterns are similar to one another and consistent with Gaussian noise. We conclude that this method did not identify any gravitational wave emitting candidates.

\subsubsection{Power-Law Dependence}\label{sec:power_law} 
A second threshold test was conducted which includes a power-law dependence. We follow the same framework presented in previous section except that the z-score threshold changes for the harmonic we are searching on. The power law used is such that the z-score$_{\rm{threshold}} \propto n^{-1}$. To generate the thresholds, we define a combined probability as in Equation~\ref{eqn:combinedprob}. We define the z-score threshold for the fundamental frequency as z-score$_{\rm{\,T}}$ which the rest of the harmonics thresholds are derived from. The first harmonic has a z-score threshold equal to z-score$_{\rm{\,T}}$/2   and the second harmonic has a z-score equal to z-score$_{\rm{\,T}}$/3. z-score$_{\rm{\,T}}$ is chosen such that the product of the associated one-tailed probability for each z-score threshold is equal the combined probability.

With this framework, we generate z-score thresholds for all sets of harmonics ranging from n = 4 to n = 23. We ran our tests for various combinations of z-score threshold that had the equivalent $\rm{Probability}_{\rm{combined}}$ values set to 3.2 $\times$ 10$^{-5}$, 2.9 $\times 10^{-7}$, 9.9 $\times 10^{-10}$, 1.3 $\times 10^{-12}$, 5.6 $\times 10^{-16}$, which corresponds to 4, 5, 6, 7 and 8$\sigma$ of a Gaussian distribution. We found that there were no fundamental frequencies and harmonics where all were above the z-score thresholds.  Therefore, we conclude that including a power-law dependence of n$^{-1}$ into the constant threshold test did not identify any gravitational wave emitting candidates.


\subsection{Harmonic Search: Unconstrained}\label{sec:harmonic_unconstrained}

In Section~\ref{sec:harmonic_constant}, we did not identify any candidate frequency bins that would be consistent with a gravitational wave emitting source. One possibility is that
while there may be a gravitational wave signal, one of the harmonic frequency bins might have an additional noise contribution. This would not meet our threshold criteria and be excluded from follow-up. 
Another possibility is that a harmonic source could emit with a power law that is very different from the $n^{-1}$ dependence we searched in Section~\ref{sec:power_law}.

Throughout this unconstrained search, we follow the steps outlined in Section~\ref{sec:algorithm}. This method identifies sets of fundamental frequencies and harmonics that could have any per-bin z-score value as long as the combined probability value is highly significant.

For each fundamental frequency, we first determine the maximum number of harmonics that would fit within the 25 MHz range. For example, for a fundamental of 2 MHz, the combined probability is calculated for 11 harmonics since the 12th harmonic would be at 26 MHz. Next, we mapped the z-scores of each frequency in this set to its corresponding probability as described in Section~\ref{sec:algorithm}. Finally, these probabilities of the fundamental frequency and $n$-harmonics were then multiplied together to get a combined probability.

After the combined probability of all fundamental frequency and harmonic sets is calculated, we selected the 15 lowest combined probability fundamental frequency sets for follow-up. Additionally, we ran this same test on the other interferometer quadrants (negative-real, positive-imaginary, negative-imaginary) and on all four quadrants of the lightbulb dataset. 

Figure~\ref{fig:harmonicTests} shows the z-score as a function of frequency and highlights the three fundamental frequencies and their harmonics with the lowest combined probability values. The harmonic emission pattern of each set does not follow a power-law dependence nor does it seem to have excess power in most bins. Furthermore, we did not identify any unique pattern beyond what was seen in the candidates identified in the negative-real, positive-imaginary, negative- imaginary interferometer quadrants and all four quadrants of the lightbulb dataset. We conclude that this method did not identify any gravitational wave emitting candidates.


\begin{figure*}
	\centering \includegraphics[width= 0.82\textwidth]{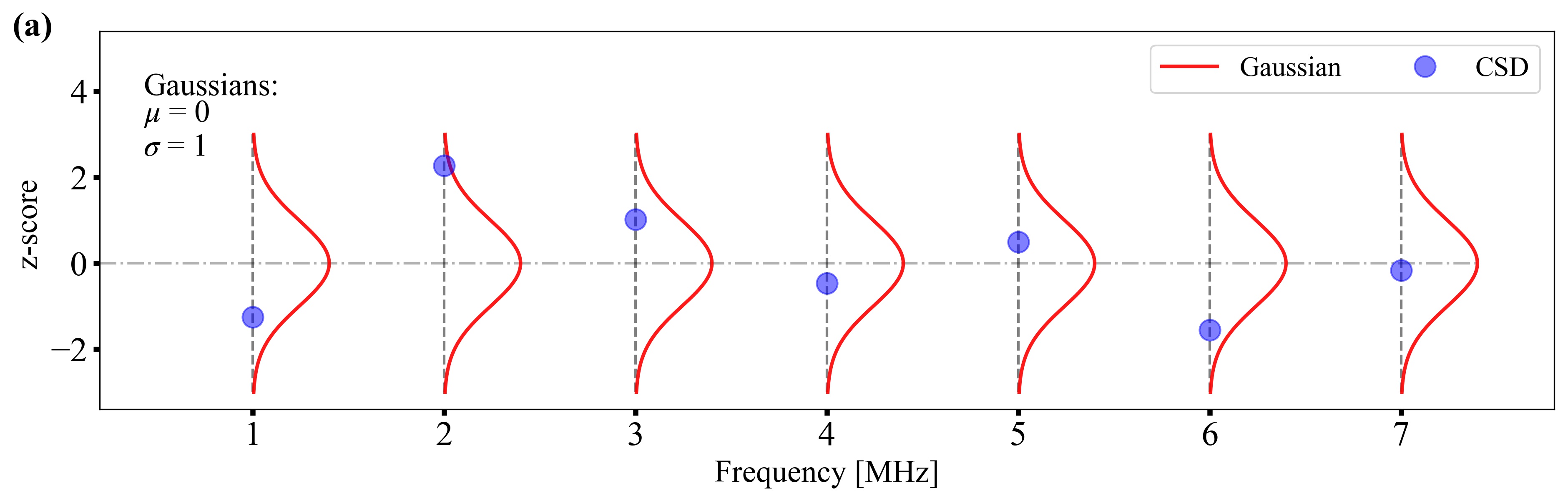} 
	\centering 
	\includegraphics[width=0.82\textwidth]{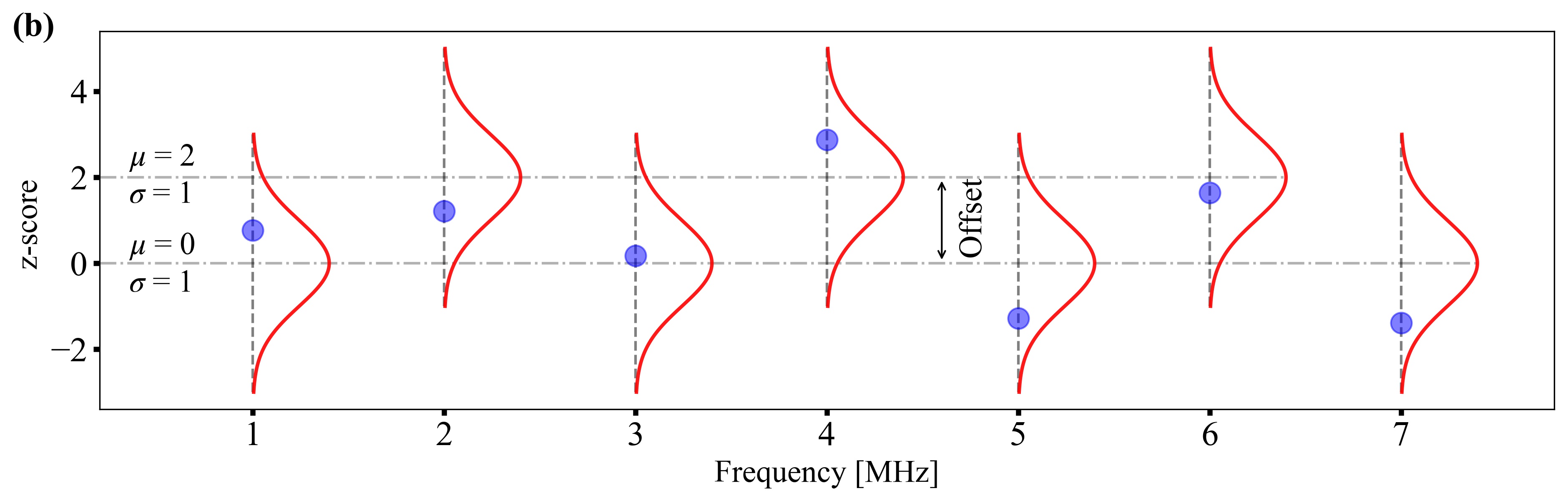}
	\includegraphics[width=0.82\textwidth]{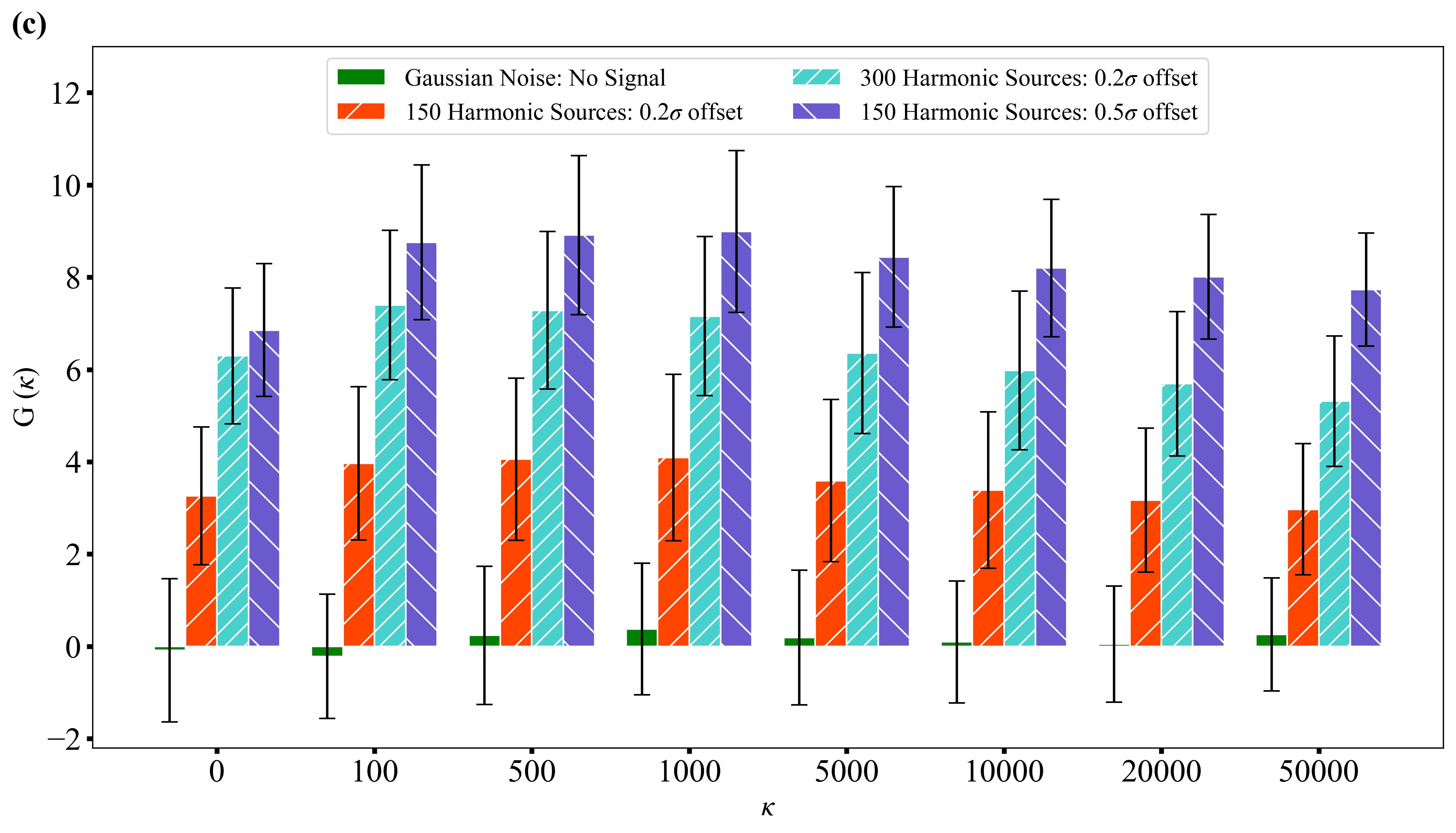}
	\caption{{To evaluate if a population of harmonic sources whose collective emission would be identifiable in the interferometer dataset, we generate simulated datasets based on an underlying Gaussian distribution of noise and populations of sources. Figure (a) is a simplified representation of Gaussian distributed noise for 7 frequency bins; whereas the full simulated dataset matches the interferometer dataset of 62,914 frequency bins. The blue dots are CSDs generated from the underlying Gaussian distributions (in red). Figure (b) is an illustration of a single injected harmonic source, which would appear as an offset in the underlying Gaussian distribution that is proportional to the power emission strength of the source. This artificial harmonic source has a fundamental frequency of 2 MHz and has two harmonics at 4 and 6 MHz. Using this framework, we generated four different simulated datasets. The first one contained pure Gaussian noise: all of the z-scores in the frequency bins were generated out of Gaussians with $\mu$ = 0. The other three datasets contained pure Gaussian noise along with injected harmonic sources: one dataset contained 150 sources with 0.2$\sigma$ offsets, the second one had 300 sources with 0.2$\sigma$, and the third one had 150 sources with 0.5$\sigma$ offsets. To characterize a population of harmonic sources, we defined a function G($\kappa$) in Section~\ref{sec:harmonic_population} which can detect whether there is a population of harmonic sources that emit in the positive-real quadrant. The $\kappa$ in the x-axis is used as a cutoff for fundamental frequencies and harmonics. Figure (c) highlights the results of the four simulated datasets ran through the population search algorithm. We see that the algorithm is sensitive to both the number of harmonic sources and their emission power. The results from this figure are used as a baseline to compare the results from the interferometer dataset to.}}
		  
	 \label{fig:simulated_datasets}
	
\end{figure*}

\subsection{Harmonic Search: Population}\label{sec:harmonic_population}

To identify individual harmonic sources, the source should have strong enough gravitational wave power that can be distinguished from statistical noise. This was the motivation behind the tests in Sections~\ref{sec:excesspower}, ~\ref{sec:harmonic_constant}, and ~\ref{sec:harmonic_unconstrained}, which did not find evidence of individual harmonic sources. As a final test, we conducted a search for a population of harmonic sources with low gravitational wave power emission whose collective emission would be identifiable.

The framework is similar to that in the unconstrained search of Section~\ref{sec:harmonic_unconstrained}. The main difference is that we use a different statistic for the population, $\rm S({P}_{\rm{mixed}})$, which is defined as 
\[\tag{4}
S(\rm{P}_{\rm{mixed}}) =
\begin{cases}
\rm{P}_{\rm{mixed}} - 1  & \text{if P$_{\rm{mixed}}$ $\geq$ 1} \\
-1 \times (\frac{1}{\rm{P}_{\rm{mixed}}} - 1) & \text{if P$_{\rm{mixed}}$ $\textless$ 1} 
\end{cases} \label{eqn:sprob}
\]

\begin{equation}
\tag{5}
\rm{P}_{\rm{mixed}} =  \prod_{i = 0}^{n}
(\rm{Probability}_{\,i})^{\,j}
\end{equation}

\[
j =
\begin{cases}
-1 & \text{if z-score$_{\rm{i}}$ $\geq$ 0} \\
1 & \text{if z-score$_{\rm{i}}$ $\textless$ 0} 
\end{cases}
\]

P$_{\rm{mixed}}$ is calculated for each fundamental frequency (and harmonics) where i is the ith harmonic of that fundamental frequency, z-score$_{\rm{i}}$ is the z-score for the ith harmonic, and the Probability$_{\rm{i}}$ is the two-tailed probability associated with its z-score; in this section, a two tailed test was used so that a z-score of 0 corresponded with a probability of 1. The exponent j is either 1 or -1 depending if the z-score for the ith harmonic is positive or negative; if z-score$_{\rm{i}}$ is positive, we divide by its corresponding probability and if negative, we multiply by its probability. This allows us to create a function which becomes symmetric by taking the inverse of P$_{\rm{mixed}}$ as seen in Equation~\ref{eqn:sprob}.

The function S(P$_{\rm{mixed}}$) is generated to detect if the z-scores of a fundamental frequency and harmonics are, on average, weighted towards the positive or negative z-score quadrant. The function S(P$_{\rm{mixed}}$) is symmetric and centered about zero.  

 As an example, if a fundamental frequency and 5 harmonics all have z-scores of 1 (corresponds with a two-tailed probability of 0.3173), S(P$_{\rm{mixed}}$) equals a large positive number: 979. If half of the z-scores are +1 and the other half are -1, S(P$_{\rm{mixed}}$) = 0. If all of the z-scores are -1, S(P$_{\rm{mixed}}$) = -979. Given that harmonic sources emit a gravitational wave signal on the positive-real quadrant, we use this function to search for a surplus of fundamental frequencies and harmonics with a positive S(P$_{\rm{mixed}}$). 
 
 To identify a population of harmonic sources, we generate S(P$_{\rm{mixed}}$) for each fundamental frequency and harmonics. Next, we assign various cutoffs, denoted as $\kappa$, and count the number of fundamental frequencies whose S(P$_{\rm{mixed}}$) are above $\kappa$ and below -$\kappa$. For example, by setting $\kappa$ = 100, we count the number of fundamental frequencies and harmonics whose S(P$_{\rm{mixed}}$) $\textgreater$ 100 and the number whose $S(P_{\rm{mixed}}) \textless -100$. By comparing the number of fundamental frequencies and harmonics above and below this cutoff, we become sensitive to a population of harmonic sources with emitted power on the positive or negative CSD quadrant.

In order to make the comparison between multiple datasets and $\kappa$ values, we create a normalized function that encapsulates whether we have a surplus of harmonic sources emitting in the positive-real quadrant:

\begin{equation}
\tag{6}
G(\kappa) =  
\frac{\rm{a}(\kappa) - \rm{b}(-\kappa)} {\sqrt{\rm{a}(\kappa) + \rm{b}(-\kappa)}} \label{eqn:g_kappa}
\end{equation}

where $\kappa$ is the cutoff for S(P$_{\rm{mixed}}$), a($\kappa$) is the number of fundamental frequencies (and harmonics) whose S(P$_{\rm{mixed}}$) is greater than $\kappa$, and b(-$\kappa$) is the number of fundamental frequencies whose S(P$_{\rm{mixed}}$) is less than -$\kappa$. The denominator normalizes the function across various $\kappa$ values.

A summary of the algorithm to test for a population of harmonic sources is as follows:

\begin{enumerate}
	\item Calculate the z-score for each frequency bin using Equation~\ref{eqn:zscore}.
	\item Assign  the  corresponding  probability  to  each  z-score values for a Gaussian distribution.
	\item Start at the first frequency in the array and assign it as a fundamental frequency. 
	\item Find the largest number of harmonics, $n$, of that fundamental frequency that can fit within 25 MHz.
	\item Calculate the S(P$_{\rm{mixed}}$) function for that fundamental frequency and its harmonics using Equation~\ref{eqn:sprob}. 
	\item Move to the next fundamental frequency and repeat steps 4 \& 5 through the entire list of fundamental frequencies. 
	\item Define cutoffs, $\kappa$, for $\pm$ S(P$_{\rm{mixed}}$).
	\item Calculate G($\kappa)$, which quantifies the difference between the number of fundamental frequencies and their harmonics whose S(P$_{\rm{mixed}}$) falls above or below $\pm$ $\kappa$.
\end{enumerate}

To measure the effectiveness of this algorithm to detect a population of harmonic sources, we created simulated datasets to run through our algorithm. First, a simulated dataset of pure Gaussian noise is generated with $\mu = 0$ and $\sigma = 1$. Figure 4a is an illustration of what the simulated dataset would look like where the blue points are the CSD z-scores for 7 frequency bins. The red lines are the underlying Gaussian statistics where the z-scores are generated from. Second, we create another simulated dataset that has pure Gaussian noise and an underlying harmonic source. A harmonic source is modeled by  adding an excess power offset to the Gaussian distribution. Figure 4b is an illustration of a simulated harmonic source that has a fundamental frequency of 2 MHz and two harmonics at 4 and 6 MHz. The simulated source would emit gravitational waves at all of these frequencies and produce an excess power of 2$\sigma$ above the noise. 

With this framework, we generate four simulated datasets to quantify the sensitivity of G($\kappa$) to a population of low signal-to-noise harmonic sources: 1) pure Gaussian noise with $\mu$ = 0 and $\sigma$ = 1, 2) Gaussian noise and 150 injected harmonic sources with 0.2$\sigma$ offsets, 3) Gaussian noise and 300 injected harmonic sources with 0.2$\sigma$ offsets, and 4) Gaussian noise and 150 injected harmonic sources with 0.5$\sigma$ offsets. The simulated datasets contain the same number of frequency bins as the interferometer dataset.

These four datasets are run through the algorithm and we plot G($\kappa$) for various $\kappa$ values in Figure 4c. For each simulated dataset that has injected harmonic sources, G($\kappa$) is a positive number that is proportional to both the number of harmonic sources and their emission power. G($\kappa$ = 0) includes every fundamental frequency and is most sensitive to a large number of harmonic sources that can have very low signal-to-noise. With $\kappa$ = 50,000, the only fundamental frequencies that are beyond this cutoff must have strong power emission offsets. Therefore, G($\kappa$ = 50,000) is primarily sensitive to harmonic sources with high signal-to-noise but can identify a smaller population of sources. Figure 4c includes a wide spread of $\kappa$ values to consider various possible scenarios of harmonic source populations. This figure validates that the G($\kappa$) function would be able to recover an underlying population of harmonic sources.

\begin{figure*}
  \centering 
  {\includegraphics[width=\textwidth]{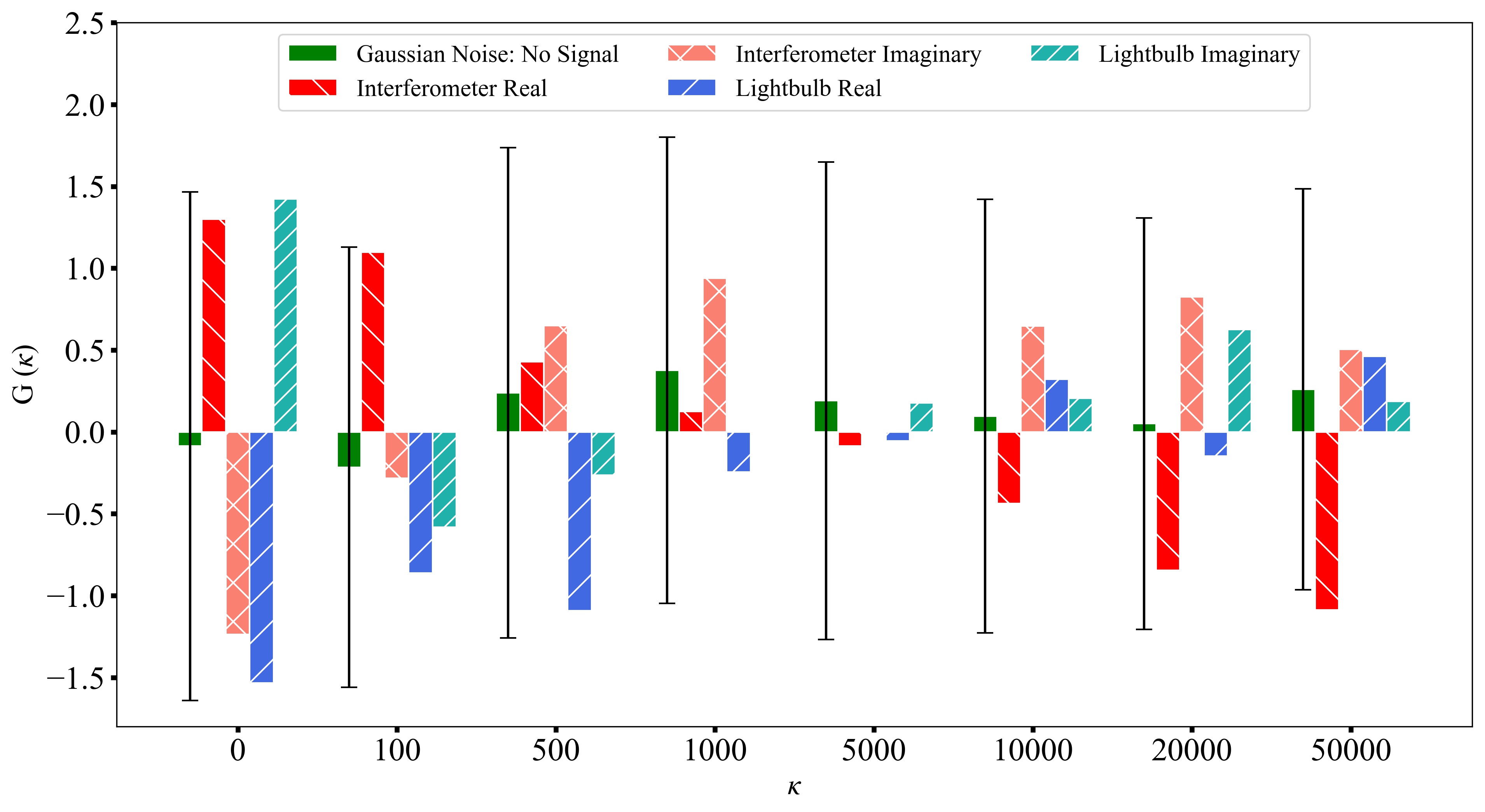}}
  \caption{Results of the search for a population of harmonic sources in the real and imaginary components of the interferometer and lightbulb datasets. G($\kappa$), defined in Section~\ref{sec:harmonic_population}, is a function whose sensitivity in detecting a population of harmonic sources is illustrated in Figure~\ref{fig:simulated_datasets}. The x-axis represents the cutoffs, $\kappa$, which is  sensitive to a wide range of population types. Low $\kappa$ values are sensitive to a large population of sources with low excess power whereas high $\kappa$ are sensitive to a smaller population with higher excess power. We include the results of the simulated data of 1$\sigma$ Gaussian noise (green) and its associated error bars generated by running 30 realizations of Gaussian noise datasets. If there were an underlying population of harmonic sources, those would appear above the errors bars of this Gaussian distribution and follow trends highlighted in Figure~\ref{fig:simulated_datasets}. We find that the interferometer real dataset (red) is consistent with pure Gaussian distributed noise. Moreover, we find that the interferometer-real results follow similar trends as the results from interferometer-imaginary (pink), lightbulb-real (blue), and lightbulb-imaginary (teal) and find that all of them are consistent with Gaussian distributed noise. We conclude there is no evidence of a population of harmonic sources across our frequency range.}\label{fig:smax_laser_light}
\end{figure*}

Given the effectiveness of G($\kappa$) to identify an underlying population of harmonic sources, we run the algorithm on the interferometer dataset. In Figure~\ref{fig:smax_laser_light}, we compare the simulated pure Gaussian noise (green) to the real component of the interferometer dataset (red). Additionally, we test the interferometer-imaginary (pink), lightbulb-real (blue), and lightbulb-imaginary (teal) for comparison.  

The results in Figure~\ref{fig:smax_laser_light} illustrate that all of the G($\kappa$) values for each dataset are consistent with the trends of 1$\sigma$ of the Gaussian noise simulated dataset. Additionally, the results for the interferometer-real data follow the same trend to those from the interferometer-imaginary data and both lightbulb datasets. We conclude that we did not find evidence of a population of low signal harmonic sources.

\section{Conclusions}
Using the Holometer as a high frequency gravitational wave detector, we conducted the first MHz search for harmonic sources. We utilized a dataset spanning over 704 hours obtained from two co-aligned, power-recycled Michelson Interferometers. This allows a search across the 1 to 25 MHz range with 382 Hz frequency resolution as described in Section~\ref{sec:ifodataset} and seen in Figure~\ref{fig:full_ifo_data}. These measurements extend the gravitational wave spectrum beyond what is accessible with PTA, LISA, LIGO, Virgo, KAGRA, and GHz experiments \cite{2015_Advanced_LIGO, VIRGO2015, amaroseoane2017laser_LISA, kerr2020parkes, GEO600_2004, EPTA2015, PPTA2010, 2018ApJS..235...37A,  PhysRevD.88.043007, Bernard2001, 2006Cruise_protopyte, 2020graviton-magnon, 2019teraHz}.

Astrophysical sources that are predicted to emit gravitational radiation in harmonics are cosmic strings \cite{DePies2007, Peters1964} and eccentric black hole binaries \cite{Peters1964, Peters1963, Tanay_2016_eccentric, Moore_2016_eccentric, Hinder_2018_eccentric, Cao_2017_eccentric, Klein_2018_eccentric, Tiwari_2019_eccentric}. In Section~\ref{sec:analysis}, we present four different ways to identify potential frequency candidates. The first three methods search for individual harmonics while the fourth looks for an underlying population. For each test, we calculate a z-score for each frequency bin, as defined in Equation~\ref{eqn:zscore}, which is the measured cross-spectral density over the standard deviation. 

In the first search (Section~\ref{sec:excesspower}), we perform an excess power search to identify any frequency bins greater than 5$\sigma$ as highlighted in Figure~\ref{fig:ifoCSDdistributions}. We did not find any frequency bins and rule out the possibility of very loud constant emitters.

Given that the fundamental and harmonic frequencies correspond to the same source, we are able to reduce the signal-to-noise threshold for each individual frequency bin. To identify potential sources, we construct a combined probability (Equation~\ref{eqn:combinedprob}) for each fundamental frequency and its $n$-harmonics. In Section~\ref{sec:harmonic_constant} and ~\ref{sec:harmonic_unconstrained} we use this framework to search for individual harmonic source candidates. 

The second search (Section~\ref{sec:harmonic_constant}) tests the hypothesis that the z-score for a fundamental frequency and all of its harmonics must lie above a z-score threshold. We conduct two versions of this search: one where we set the same z-score threshold for all frequencies in a set of fundamental frequency and harmonics (Section~\ref{sec:constant_threshold}) and one where the z-score threshold decreases according to a power law $\propto$ n$^{-1}$ for increasing harmonics (Section~\ref{sec:power_law}). We set the combined probability value to ones that corresponds with a 4, 5, 6, 7, and 8$\sigma$ detection and compared z-score values to the corresponding minimum per-bin-threshold for harmonics ranging from $n=4$ up to $n=23$. We did not identify any candidates that are likely to be gravitational wave emitting harmonic sources in either search.

Our third search (as described in Section~\ref{sec:harmonic_unconstrained}) was designed to allow for a wider range of power law dependencies beyond $\propto$ n$^{-1}$ with increasing frequencies. Additionally, this allows for the possibility of a few contaminated frequency bins that the minimum threshold search would not identify. To do this, we remove the per-bin minimum threshold. Rather, we multiply the probabilities of each z-score for the fundamental and harmonic frequencies and arrive at a combined probability. This number is used to identify fundamental frequencies of interest. 

We searched within the positive-real component of the CSD and identified the 15 fundamental frequencies and harmonics with the lowest combined probability. Three of these identified fundamental frequencies are seen in Figure~\ref{fig:harmonicTests}. These were examined and their emission pattern was compared to those found in the other interferometer quadrants (negative-real,  positive-imaginary, negative-imaginary) and on all four quadrants of the lightbulb dataset. We did not identify any unique pattern beyond what was seen in these datasets and conclude this method did not find any gravitational wave emitting candidates. 

As a final test, we consider a population of harmonic sources emitting at low powers whose collective signal would be identifiable (as described in Section~\ref{sec:harmonic_population}). We define an equation (Equation~\ref{eqn:sprob}) that enables us to determine whether z-scores of a fundamental frequency and harmonics are weighted towards the positive or negative quadrants. We set various cutoff values on this function and count the number of fundamental frequencies and harmonics whose function values are greater than or less than the cutoffs. We test the sensitivity of this function on 4 different simulated datasets with underlying Gaussian distributions as shown in Figure~\ref{fig:simulated_datasets}. We find that the algorithm is sensitive to comparing if there is a surplus of fundamental frequencies whose z-scores are mainly in the positive z-score quadrants. 

Using Equation~\ref{eqn:g_kappa}, we search in the real component of the interferometer data for a population of harmonic sources. We compare those results to simulated Gaussian noise and find that these two datasets are consistent. Additionally, we compare the the imaginary-interferometer, real-lightbulb, and imaginary-lightbulb datasets, which can be seen in Figure~\ref{fig:smax_laser_light}. We find that the real-intereferometer dataset is consistent with these three datasets and that all datasets have values consistent with Gaussian distributed noise. This method did not identify any trends that indicate an underlying population of harmonic sources. 

Our analysis choices allowed us to explore deeper into the Holometer data and search for potential gravitational wave candidates within the 1-25 MHz frequency range. Based on these four analysis methods, we conclude that there are no identifiable harmonic sources. In the broader context of placing constraints on cosmic string loops and eccentric black hole binaries, these experimental results should motivate further theoretical studies, which is beyond the scope of this paper.

\section{Acknowledgments}
We first want to acknowledge the support for the Holometer instrument and collaboration itself which made it possible to preform this analysis. The Holometer was supported through the Department of Energy at Fermilab under Contract No. DE-AC02-07CH11359 and the Early Career Research Program (FNAL FWP 11-03), and by grants from the John Templeton Foundation, the National Science Foundation (Grants No. PHY-1205254 and No. DGE-1144082), NASA (Grant No. NNX09AR38G), the Fermi Research Alliance, the Kavli Institute for Cosmological Physics, University of Chicago/Fermilab Strategic Collaborative Initiatives, and the Universities Research Association Visiting Scholars Program. Second, we would like to acknowledge support that enabled our research endeavors. Jeronimo Martinez  Garcia-Cors was in part supported by the Kavli Institute of Cosmological Physics. Brittany Kamai was supported by Caltech California Alliance, Fisk-Vanderbilt Bridge and the Heising-Simons Postdoctoral Fellowships. Third, we would like to acknowledge the incredible mentorship provided to us by Stephen S. Meyer and Craig Hogan. Both of them provided support and guidance throughout this entire project and more broadly had a strong influence on who we are as scientists. Fourth, we'd like to gratefully acknowledge the fruitful discussions with many talented scientists such as Sotiris Sandias, Katelyn Brevik, Kelly Holley-Bockelmann, Jess McIver, Lee McCuller and Jonathan Richardson about the theoretical and experimental scope of this project. Fifth, we would like to acknowledge the Indigenous people whose ancestral lands we were on during this project: we'd like to send our gratitude to the Kiikaapoi, Peoria, Bod$\rm{\text{\`{e}}}$wadmiakiwen, Miami and O$\hat{\rm{c}}$eti $\hat{\rm{S}}$akwin, Tongva, and Chumash peoples (thanks to the Native Land Digital project for making this  information accessible). Sixth, we would like to extend our sincerest gratitude to our anonymous CQG reviewer whose detailed feedback helped shape this paper in substantial ways. Lastly, we would like to highlight the unconditional support we had from our friends and families
throughout the complexities associated with completing a new research project.

\bibliographystyle{ieeetr}
\bibliography{./MHzCosmicStrings_unique} 
\end{document}